\documentclass[twocolumn]{aastex631}
\usepackage{amsmath}
\usepackage{appendix}
\usepackage{float}
\usepackage{multirow}
\usepackage{hyperref}
\usepackage{nameref}

\begin{document}

%\svgsetup{inkscapelatex=false}

\newcommand{\Fp}[1]{\textcolor{red}{#1}}

\newcommand{\bs}[1]{%
\boldsymbol{#1}
}%
\newcommand{\bA}{\textit{\textbf{A}}}
\newcommand{\bB}{\textit{\textbf{B}}}
\newcommand{\bU}{\textit{\textbf{U}}}
\newcommand{\ba}{\textit{\textbf{a}}}
\newcommand{\bb}{\textit{\textbf{b}}}
\newcommand{\be}{\textit{\textbf{e}}}
\newcommand{\bE}{\textit{\textbf{E}}}
\newcommand{\bh}{\textit{\textbf{h}}}
\newcommand{\bH}{\textit{\textbf{H}}}
\newcommand{\bJ}{\textit{\textbf{J}}}
\newcommand{\bj}{\textit{\textbf{j}}}
\newcommand{\bk}{\textit{\textbf{k}}}
\newcommand{\bp}{\textit{\textbf{p}}}
\newcommand{\bq}{\textit{\textbf{q}}}
\newcommand{\br}{\textit{\textbf{r}}}
\newcommand{\bu}{\textit{\textbf{u}}}
\newcommand{\bx}{\textit{\textbf{x}}}
\newcommand{\by}{\textit{\textbf{y}}}
\newcommand{\bz}{\textit{\textbf{z}}}
\newcommand{\bw}{{\boldsymbol{\omega}}}
\newcommand{\dd}{\mathrm{d}}
\newcommand{\ii}{\mathrm{i}}
\newcommand{\ee}{\mathrm{e}}
\newcommand\Rm{\textit{Rm}}
\newcommand{\sign}{\text{sign}}

\title{Can the dynamo of spiral-arm galaxies be explained by anisotropic conductivity ?}

\author{Paul Gomez}
\affiliation{Université Lyon 1, ENS de Lyon, CNRS, Laboratoire de Géologie de Lyon, Lyon 69622, France; \href{mailto:[paul.gomez@univ-lyon1.fr]}{paul.gomez@univ-lyon1.fr}, \href{mailto:[thierry.alboussiere@ens-lyon.fr]}{thierry.alboussiere@ens-lyon.fr}}

\author{Franck Plunian}
\affiliation{Université Grenoble Alpes, University of Savoie Mont Blanc, CNRS, IRD, Université Gustave Eiffel, ISTerre, 38000 Grenoble, France; \href{mailto:[franck.plunian@univ-grenoble-alpes.fr]}{franck.plunian@univ-grenoble-alpes.fr}}

\author{Thierry Alboussière}
\affiliation{Université Lyon 1, ENS de Lyon, CNRS, Laboratoire de Géologie de Lyon, Lyon 69622, France; \href{mailto:[paul.gomez@univ-lyon1.fr]}{paul.gomez@univ-lyon1.fr}, \href{mailto:[thierry.alboussiere@ens-lyon.fr]}{thierry.alboussiere@ens-lyon.fr}}

%\linenumbers

%%%%%%%%%%%%%%%%%
\begin{abstract}
%%%%%%%%%%%%%%%%%

The possibility of generating a magnetic field by dynamo effect with anisotropic electrical conductivity rather than turbulent flow has been demonstrated theoretically \citep{plunian2020axisymmetric} and experimentally \citep{alboussiere2022fury}. If the electrical conductivity is anisotropic, the electrical currents will flow preferentially in certain directions rather than others, and a simple differential rotation will suffice to generate a magnetic field. In a galaxy with spiral arms, it is reasonable to assume that the electrical conductivity will be twice larger along the arms than in the perpendicular direction, suggesting the possibility of an anisotropic dynamo.
However, a further geometrical criterion must be satisfied to obtain a dynamo \citep{plunian_fast_2022}. It is given by $\Omega' \cdot\sin p  > 0$, where $p$ is the pitch angle of the spiral arms, with $p \in[-\frac{\pi}{2}, \frac{\pi}{2}]$, and $\Omega'$ is the radial shear of the angular velocity.
We find that the usual spiral arms galaxies, which satisfy $|\Omega'|<0$, do not satisfy this dynamo condition because they have trailing arms instead of leading arms. Even the galaxy NGC 4622, which has both trailing and leading arms, does not satisfy this dynamo condition either. This is confirmed by numerical simulations of the induction equation. Thus, for all the spiral arms galaxies known to date, the anisotropy of the spiral arms cannot explain the existence of galactic magnetic fields until further notice.
\end{abstract}


%%%%%%%%%%%%%%%%%%%%%%%
\section{Introduction}
%%%%%%%%%%%%%%%%%%%%%%%
Galactic magnetic fields have been observed for decades, raising questions about their role in the formation and behaviour of galaxies. They are composed of a large-scale coherent part and a small-scale random part due to turbulence. A typical time-scale for turbulent magnetic diffusion in galaxies is $10^8$ years. This corresponds to about one tenth of the age of a galaxy, implying that a physical mechanism is needed to explain why galactic magnetic fields are still present. Assuming that a galaxy can be modelled by a rotating plasma, the dynamo effect is a good candidate to explain the generation of galactic magnetic fields \citep{brandenburg2005astrophysical, beck_synthesizing_2019}.
Since the pioneering work of \cite{krauseMHD1981}, an alpha-effect due to small-scale helical turbulence, conjugated to a strong differential rotation, has been proposed as the main dynamo mechanism.

The magnetic Reynolds number is defined by 
\begin{equation}
    R_m = \sigma\mu U L,
    \label{eq:Rm}
\end{equation}
where $\sigma$ is the electrical conductivity of the fluid, $\mu$ its magnetic permeability, $U$ and $L$ two characteristic values of velocity and length scale. This dimensionless number can be understood as the ratio of the magnetic diffusion time by the flow turn over time, implying that a necessary condition for dynamo action is $R_m > 1$.  In galaxies $R_m \approx 10^{18}-10^{20}$ \citep{schekochihin2002spectra, brandenburg2005astrophysical}, much higher than that required for dynamo action. However, a question arises: is the dynamo still efficient at high $R_m$ or, in other words, is the dynamo of fast type \citep{childress2008stretch}. A fast dynamo corresponds to a strictly positive kinematic growth rate of the magnetic field, as opposed to a slow dynamo for which the magnetic growth rate vanishes in the limit of a high $R_m$. The turbulent alpha-effect is suspected to quench at high $R_m$ due to non-linear effects, challenging the usual dynamo mechanism advocated to explain the galactic magnetic field. \citep{rincon2019dynamo}.

Another possibility for generating a magnetic field by dynamo effect has recently been  proposed \citep{alboussiere2020dynamo,plunian2020axisymmetric}. It is based on large-scale differential rotation and anisotropic electrical conductivity. Turbulence is not necessary for this dynamo mechanism to work.
Assuming complete axisymmetry, it has been shown that an anisotropic dynamo can occur provided the differential rotation is higher than some threshold depending on the properties of the anisotropy \citep{plunian2020axisymmetric}. 
For an angular velocity with a smooth radial profile, an asymptotic approach has shown that the anisotropic dynamo is fast \citep{plunian_fast_2022}, then proving to be an interesting candidate for a galactic dynamo. The anisotropic dynamo action has also been demonstrated experimentally \citep{alboussiere2022fury}. Although the geometry of this experiment is different from that of a galaxy, they both share a logarithmic spiral anisotropy of electrical conductivity.

In this paper, we study the possibility for a spiral arms galaxy to work as an anisotropic dynamo. 
In section \ref{sec:modeling}, the modeling of the anisotropic dynamo is explained. In section \ref{sec:simulations}, some results of simulations are shown, for several rotation profiles and several arms configurations, trailing, leading arms or a combination of both.

%%%%%%%%%%%%%%%%%%%%%%%
\section{Modeling}
\label{sec:modeling}
%%%%%%%%%%%%%%%%%%%%%%%%

%%%%%%%%%%%%%%%%%%%%%%%%
\subsection{The pitch angle $p$}
\begin{figure}%[t]
    \includegraphics[width=0.9\linewidth]{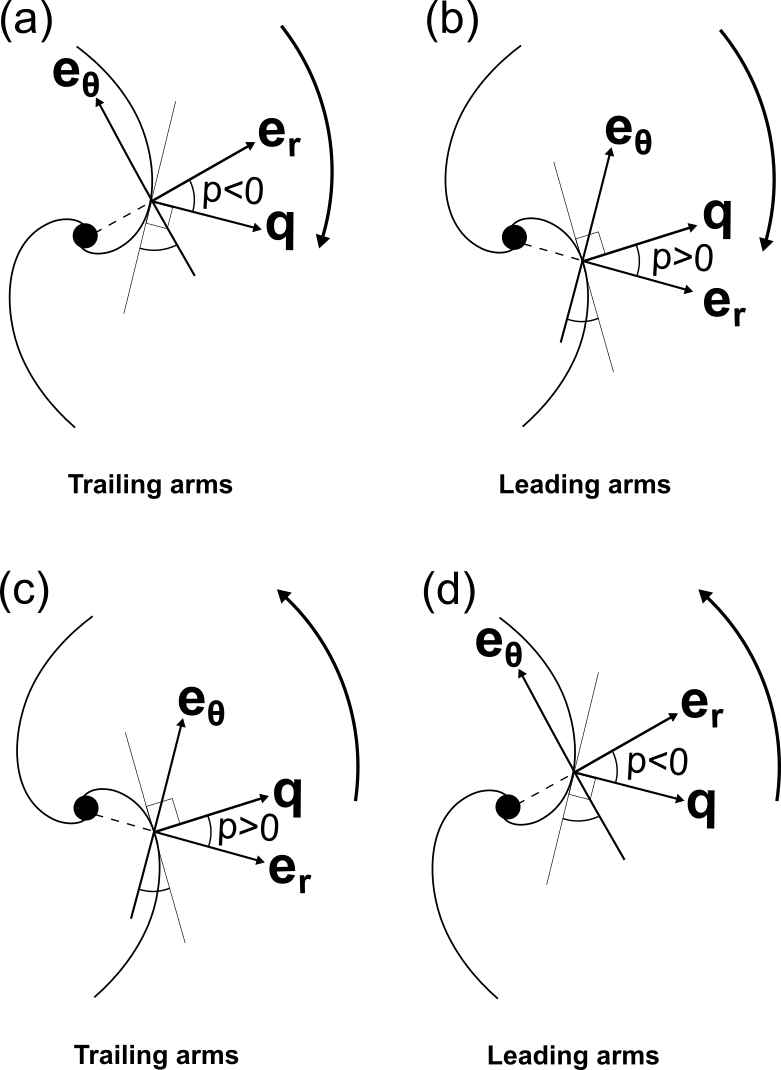}
    \centering
    \caption{Examples of leading and trailing arms, with positive or negative pitch angles $p$.
 In (a) and (b) the galaxy rotates clockwise. In (d) and (c) the galaxy rotates counterclockwise.
 %\Fp{Ca me semblerait plus claire de mettre l'arc de cercle de l'autre côté (entre $\be_{\theta}$ et la direction tangentielle à la spirale.}
 }
    \label{fig:spiral}
\end{figure}

%%%%%%%%%%%%%%%%%%%%%%%%
The shape of galactic spiral arms can be approximated by logarithmic spirals \citep{seigar1998structure} with a constant pitch angle $p$, although in reality this can vary by a few degrees \citep{savchenko2013pitch}.  
Following \cite{davis2012measurement}, the absolute value of $p$ is defined as the smallest angle between the azimuthal direction and the tangential direction of the spiral, at a given radius.
%radial direction and the direction perpendicular to the spiral, at a specified radius. 
In addition, the sign of $p$ defines the chirality of the spiral. A clockwise outwards or anticlockwise inwards winding corresponds to $p>0$, a clockwise inwards or anticlockwise outwards winding to $p<0$, as illustrated in Fig.\ref{fig:spiral}.

Denoted by $\bs{q}$, the unit vector normal to the spiral is defined by
\begin{equation}
    \bs{q} = \cos p\;\bs{e}_r + \sin p\;\bs{e}_\theta.
\end{equation}
with $p\in [-\frac{\pi}{2},\frac{\pi}{2}]$, and where $(\be_r,\be_{\theta},\be_z)$ is the cylindrical coordinate system.
%\footnote{The definition of $\bs{q}$ and $p$ are identical to the ones in \cite{plunian_fast_2022}.}
Depending on the chirality and the pitch angle, the galaxies arms are either trailing arms or leading arms as illustrated in Fig. \ref{fig:spiral}.
A leading arm has its outer tip pointing in the direction of the galaxy's rotation, whereas the outer tip of a trailing arm points in the opposite direction of rotation.

%%%%%%%%%%%%%%%%%%%%%%%%%%%%%%%%%%%%%%%%%%%%%%%%%
\subsection{Anisotropic electrical conductivity}
%%%%%%%%%%%%%%%%%%%%%%%%%%%%%%%%%%%%%%%%%%%%%%%%%

In galaxies, the electron density is greater in the spiral arms than in the interarm regions by a factor two \citep{gutierrez2010galaxy}. However, the Spitzer resistivity does not depend on the electron density \citep{spitzer1953transport} and then cannot lead to an anisotropic electrical conductivity. A candidate for such anisotropy can be the different abundance of HII regions within the spiral arm and inter-arm regions \citep{knapen1998statistical}. Another option could be to rely on the turbulent diffusivity which is most effective in the spiral arms.

We define an anisotropic effective electrical conductivity as a tensor \citep{ruderman1984magnetic},
\begin{equation}
\lbrack\sigma_{ij}\rbrack=\sigma_{\perp} \delta_{ij} + (\sigma_{\parallel}-\sigma_{\perp})q_i q_j,
\label{eq:conductivity permeability tensors}
\end{equation}
where $\sigma_{\parallel}$ and $\sigma_{\perp}$ are the electric conductivity in the directions parallel and perpendicular to $\bs{q}$.
%. In the numerical simulations we take a ratio 
We do not have an estimate for the ratio $\sigma_{\perp} / \sigma_{\parallel}$, but the key point is that it must be strictly larger than unity. For the numerical applications, we choose  $\sigma_{\perp} / \sigma_{\parallel} = 2$.

We note that the conductivity model given by (\ref{eq:conductivity permeability tensors}) implies that the conductivity is axisymmetric, which is obviously not the case, since the spiral arms break this axisymmetry. It is, however, the simplest and probably the most dynamo capable model we can think of, for studying the possibility of an anisotropic dynamo in galaxies. 
%%%%%%%%%%%%%%%%%%%%%%%%%%%%%%%%%%%%%%%%%%%%%%%%%
\subsection{Induction equation}
%%%%%%%%%%%%%%%%%%%%%%%%%%%%%%%%%%%%%%%%%%%%%%%%%
The induction equation can be written as
\begin{equation}
\partial_t \bB = \nabla \times (\bU \times \bB) - \mu_0^{-1}\nabla \times \left( \lbrack\sigma_{ij}\rbrack^{-1} \nabla \times \bB \right) ,
\label{eq:induction equation}
\end{equation}
where $\bB(r,t)$ is the magnetic induction, $\bU=U(r)\be_{\theta}$, and $\mu_0$ the vacuum magnetic permeability.

Since the velocity is stationary, and assuming axisymmetry, we  look for a magnetic induction in the form
\begin{equation}
\bB=\bB(r,z)\exp(\gamma t),
\label{eq:magnetic induction}
\end{equation}
leading to the resolution of a 2D eigenvalue problem.
In (\ref{eq:magnetic induction}), $\gamma>0$ is the signature of dynamo action, the dynamo threshold corresponding to $\gamma=0$.

%%%%%%%%%%%%%%%%%%%%%%%%%%%%%%%%%%%%%%%%%%%%%%%%%%%
\subsection{A necessary condition for dynamo action}
%%%%%%%%%%%%%%%%%%%%%%%%%%%%%%%%%%%%%%%%%%%%%%%%%%%%

In \citet{plunian_fast_2022}, an asymptotic expression of the magnetic growthrate $\gamma$ for $R_m\gg1$ and $\sigma_{\perp}\gg\sigma_{\parallel}$, has been derived,
\begin{equation}
        \gamma \approx r_0\cdot\Omega'(r_0)\cdot\cos p\cdot\sin p  - K^2,
        \label{condition}
\end{equation}
where $\Omega'(r)$ is the radial derivative of the angular velocity $\Omega(r)=U(r)/r$, $K$ a dimensionless number, and $r_0$ a value of $r$ at which the asymptotic expansion leading to (\ref{condition}) has been made. 

For $p\in [-\frac{\pi}{2},\frac{\pi}{2}]$, a necessary condition for anisotropic dynamo action is therefore that there must be at least one value of $r$ such that
\begin{equation}
    \Omega'\cdot\sin p > 0.
    \label{condition2}
\end{equation} 
Then, applying (\ref{condition2}) to cases (a-d), and depending whether the absolute value of the angular velocity $|\Omega|$ is decreasing ($|\Omega|'<0$) or increasing ($|\Omega|'>0$), the sign of $\Omega' \cdot\sin p$ can be determined (see Tab.\ref{table:signe}).
We find that if $|\Omega(r)|$ decreases (respectively increases), the dynamo is only possible in cases (b) and (d) (respectively in cases (a) and (c)).
\begin{center}
    \begin{table}[h]
    \begin{tabular}{c|c|c|c|c|c}
    &$\Omega$& $p$ &$|\Omega|'$&$\Omega'$&$\Omega'\cdot\sin p$\\
    \hline
    \multirow{2}{6em}{Trailing (a)}&\multirow{2}{0.5em}{-}&\multirow{2}{0.5em}{-}&+&-&+\\
    &&&-&+&-\\
    \multirow{2}{6em}{Leading (b)}&\multirow{2}{0.5em}{-}&\multirow{2}{1em}{+}&+&-&-\\
    &&&-&+&+\\
    \multirow{2}{6em}{Trailing (c)}&\multirow{2}{1em}{+}&\multirow{2}{1em}{+}&+&+&+\\
    &&&-&-&-\\
    \multirow{2}{6em}{Leading (d)}&\multirow{2}{1em}{+}&\multirow{2}{0.5em}{-}&+&+&-\\
    &&&-&-&+\\
    \end{tabular}
    \caption{Sign of $\Omega'\cdot\sin p$ for the cases (a-d) depicted in Fig. \ref{fig:spiral}.}
    \label{table:signe}
    \end{table} 
\end{center}
%%%%%%%%%%%%%%%%%%%%%%%%%%%%%%%%%%%%%%%%%%%%%%%%%
\subsection{Application to usual spiral arms galaxies}
\label{sec:galaxies}
%%%%%%%%%%%%%%%%%%%%%%%%%%%%%%%%%%%%%%%%%%%%%%%%%%
The central part of a galaxy, called the bulge, is made of a high concentration of stars. 
In the vast majority of spiral arms galaxies observed so far, the azimuthal velocity $U(r)$ is a solid body rotation within the bulge, and flat outside the bulge \citep{rubin1982rotational, sofue2001rotation}. This corresponds to an angular velocity $\Omega(r)$ which is constant within the bulge and whose absolute value decrases outside, corresponding to $|\Omega|'\le 0$.
From Tab.\ref{table:signe}, only the cases (b) and (d) are likely to behave as anisotropic dynamos. They both correspond to a leading arms geometry (Fig.\ref{fig:spiral}). However, all galaxies observed so far have trailing arms \citep{pasha1982direction}, so they are unlikely to be the seat of an anisotropic dynamo.

%%%%%%%%%%%%%%%%%%%%%%%%%%%%%%%%%%%%%%%%%%%%%%%%%%%%%
\subsection{NGC 4622 : an unusual spiral arms galaxy}
\label{sec:NGC 4622}
\begin{figure}%[h]
    \includegraphics[width=1.6in]{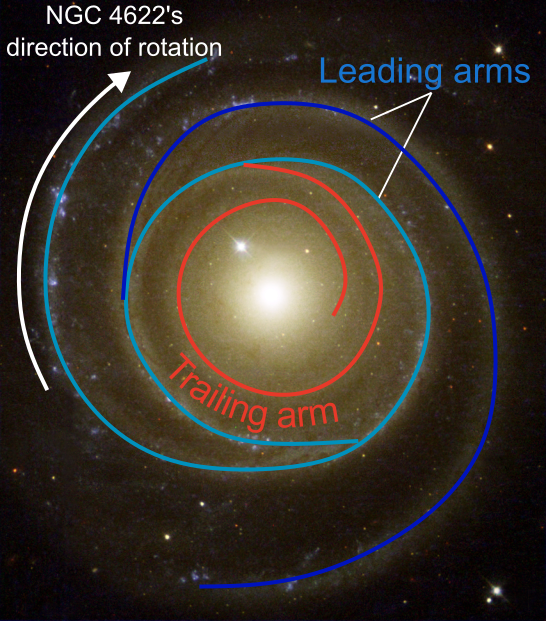}
    \includegraphics[width=1.7in]{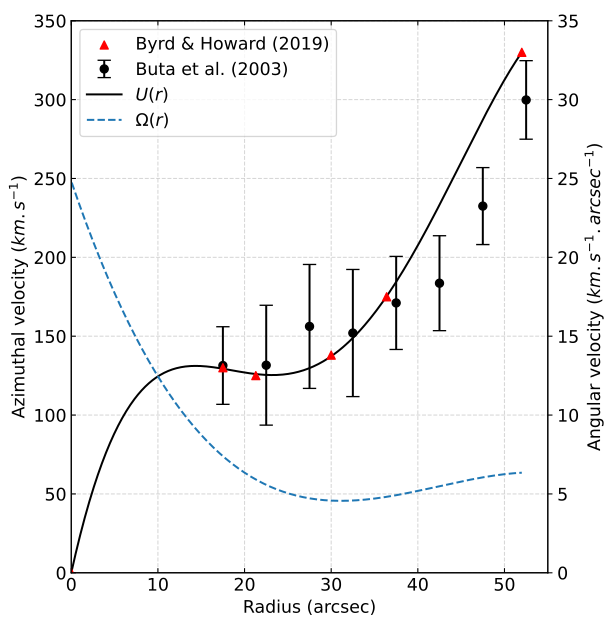}
    \centering
    \caption{(a) Left: Illustration of the trailing and leading arms of the NGC 4622 galaxy. (b) Right: The black dots and red triangles correspond to observations of the azimuthal velocity $U(r)$ by respectively \cite{buta_ringed_2003} and \cite{byrd2019ngc}. The full curve is a fit of the red triangles, corresponding to $U(r)$, and the dashed curve to $\Omega(r)=U(r)/r$.}
    \label{fig:NGC4622}
\end{figure}

%%%%%%%%%%%%%%%%%%%%%%%%%%%%%%%%%%%%%%%%%%%%%%%%%%%%%%%
The spiral arms galaxy NGC 4622 possesses one trailing arm surrounded by two leading arms \citep{byrd2019ngc}, as illustrated in Fig. \ref{fig:NGC4622}a. The trailing arm corresponds to case (a), and the two leading arms to case (b). In addition, according to Fig. \ref{fig:NGC4622}b, the azimuthal velocity is flat for the trailing arm ($|\Omega|'< 0$) and increasing in absolute value for the leading arms ($|\Omega|'> 0$), leading in both cases to $\Omega'\cdot\sin p <0$. Once again, it is therefore unlikely that the galaxy NGC 4622 is the seat of an anisotropic dynamo.

%%%%%%%%%%%%%%%%%%%%%%%%%%%%%%%%%%%%%%%%%%%%%
\section{Numerical simulations}
\label{sec:simulations}
%%%%%%%%%%%%%%%%%%%%%%%%%%%%%%%%%%%%%%%%%%%%%
In order to verify the theoretical conclusions of sections \ref{sec:galaxies} and \ref{sec:NGC 4622}, we now carry out numerical simulations.
\subsection{Velocity profiles}
%%%%%%%%%%%%%%%%%%%%%%%%
The geometry of a spiral-arm galaxy is approximated by a cylinder of height $h$ and radius $R \gg h$.
The radius of the galactic bulge is denoted $r_b$.
%The central part of the galaxy, called the bulge, which is made of a high concentration of stars, is taken of radius $r_b$. In our simulations we take 
Most spiral arms galaxies  have dimensions $r_b \sim h\approx 0.2$ to 2 kpc and $R\approx$ 10 to 50 kpc \citep{molla2000galactic, mollenhoff2004disk, zhao2004determination}. In the simulations we take $r_b = h$ and $R/r_b=50 $.
For the NGC 4622 galaxy, according to \cite{buta_ringed_2003}, $r_b\approx 8.8$ arcsec, $R\approx 52$ arcsec, and $h \sim 1.93$ arcsec (5 kpc = 29.7 arcsec). In the simulations we take $R/h=27$ and $r_b /h = 4.6$. 

We consider four azimuthal velocity profiles, $U_K$, $U_F$, $U_S$ and $U_{4622}$. The first two, $U_K$ and $U_F$, correspond to a velocity profile which is a solid body rotation within the bulge, and respectively Keplerian or flat outside the bulge. The last two, $U_S$ and $U_{4622}$, are smooth functions of $r$ closed to solid rotation for $r\ll r_b$, and (i) asymptotically flat in the limit $r\gg 1$ for $U_S$, (ii) corresponding to the observations of NGC 4622 given by the red triangles in Fig. \ref{fig:NGC4622}b for $U_{4622}$.

They are given by
\begin{eqnarray}
&U_K(r) &=\left\{
\begin{split}
    &U_0r/r_b         &\quad\quad\quad\quad\text{for}\quad   &0 \le r \le r_b \\
    &U_0\sqrt{r_b/r}  &\quad\quad\quad\quad\text{for}\quad   &r_b \le r \le R
\end{split}
\right., \label{eq:UK}\\
&U_F(r) &=\left\{
\begin{split}
&U_0r/r_b  &\quad\quad\quad\quad\quad\text{for}&\quad  0 \le r \le r_b \\
&U_0  &\quad\quad\quad\quad\quad\text{for}&\quad  r_b \le r \le R 
\end{split} \right.,\label{eq:UF}\\
&U_S(r) &=U_0\tanh(\frac{ar}{r_b}) \;\quad\quad\quad\quad\text{for}\quad 0 \le r \le R,\label{eq:US}\\
&U_{4622}(r)&= U_0\left(br+cr^2+dr^3+er^4\right) \text{for}\quad 0 \le r \le R.\quad \quad\label{eq:U4622}
\end{eqnarray}

\begin{figure}[t]
    \includegraphics[width = \linewidth]{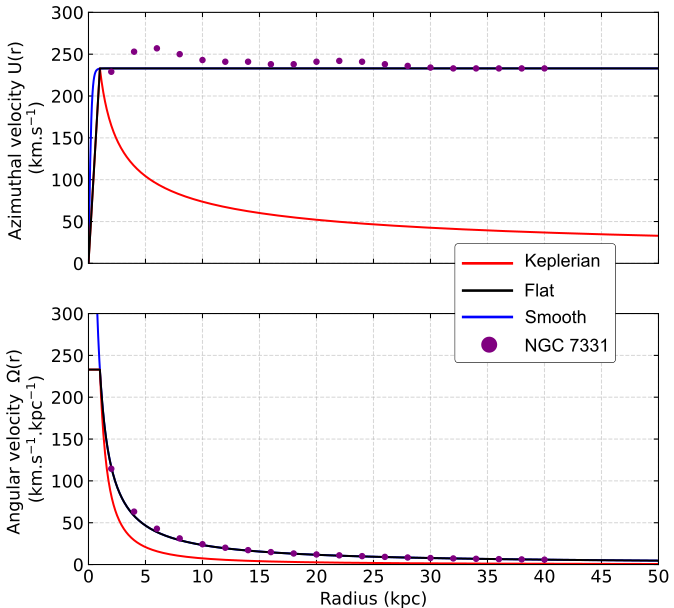}
    \centering
    \caption{(a) Top: Azimuthal velocity $U(r)$. (b) Bottom: Angular velocity $\Omega(r)=U(r)/r$. They are given for the three profiles Keplerian (K), Flat (F) and Smooth (S), and from observations of the galaxy NGC 7331 \citep{kornelis7331}. For $r\ge1$, the Flat and Smooth curves coincide.}
    \label{fig:curves}
\end{figure}

In (\ref{eq:US}), we set $U_S(r_b) = U_0$, implying that  $a=3.8$. 
The parameters $r_b$ and $U_0$ are the characteristic scale and velocity by which the induction equation (\ref{eq:induction equation}) is normalized, the problem then depending only on the magnetic Reynolds number
\begin{equation}
    R_m = \sigma_{\perp}\mu_0 U_0 r_b,
    \label{eq:Rm2}
\end{equation}
where $\mu_0$ is the magnetic permeability of void.
As an example, shown in Fig. \ref{fig:curves}a, taking $U_0=233\;$km~s$^{-1}$ and $r_b=1\;$kpc in (\ref{eq:UF}) and (\ref{eq:US}) leads to a reasonable approximation of the velocity profile measured in NGC 7331 \citep{kornelis7331}.  The corresponding angular velocity $\Omega(r)=U(r)/r$ is given in Fig. \ref{fig:curves}b. Although not represented, the value of $\Omega_S$ at $r=0$ is equal to  $885.4\;$km~s$^{-1}$~kpc$^{-1}$. 

In (\ref{eq:U4622}), the coefficients $b,c,d,e$ are derived from a 4th order polynomial regression in order to fit the data by \cite{byrd2019ngc} given by the red triangles in Fig.\ref{fig:NGC4622}b.
They are given by 
$b=7.523\times10^{-2}$~arcsec$^{-1}$, $c=-4.896\times10^{-3}$~arcsec$^{-2}$, $d=1.244\times10^{-4}$~arcsec$^{-3}$, and $e=-9.805\times10^{-7}$~arcsec$^{-4}$.
In Fig. \ref{fig:NGC4622}b, the full curve corresponds to $U(r)$ given by (\ref{eq:U4622}) with $U_0=330$~km~s$^{-1}$, and the dashed curve to $\Omega(r)=U(r)/r$. We note that $\Omega'=0$ for $r\approx 30$~arcsec \citep{byrd2019ngc}. 

%, with a velocity $U=138$ km/s
%-3.23554299e-04.x^4  4.10586249e-02.x^3 -1.61552346e+00.x^2 2.48245420e+01.x  1.55267691e-02
%  U_{4622}(r) = U_1\left(-9.805\cdot10^{-7}r^4 + 1.244\cdot10^{-4}r^3-4.896\cdot10^{-3}r^2 + 7.523\cdot10^{-2}r + 4.705\cdot10^{-5}\right) 

%%%%%%%%%%%%%%%%%%%%%%%%%%
\subsection{ Values taken for the pitch angle}
%%%%%%%%%%%%%%%%%%%%%%%%%
According to \cite{buta1992leading}, in the galaxy NGC 4622, the pitch angle is $p=-3.8^\circ$ (respectively $p=7.5^\circ$) for the part corresponding to the trailing arm (respectively leading arms). For the other spiral arms galaxies the pitch angle varies between $10^\circ$ to $50^\circ$ \citep{Seigar_2006}, and we take the mean value $p=28^\circ$.

%%%%%%%%%%%%%%%%%%%%%%%%%%%%%%%%%%%%%%%%%%%%%
\subsection{Poloidal-Toroidal decomposition}
%%%%%%%%%%%%%%%%%%%%%%%%%%%%%%%%%%%%%%%%%%%%%

%%%%%%%% Rajouter inforamtion sur paramètres et équation décomposition %%%%%%%%%%%%%%%%%%%%%%%%

The simulations are performed using Freefem++, a partial differential equation solver that employs the finite element method \citep{MR3043640}. We use a poloidal-toroidal decomposition of $\bB$, 
\begin{equation}
    \bB = \nabla\times(P(r,z)\be_\theta)+T(r,z)\be_\theta,
\end{equation}
such that the solenoidality condition $\nabla\cdot \bB=0$ is necessarily satisfied.

\begin{figure}%[h]
    \includegraphics[width = \linewidth]{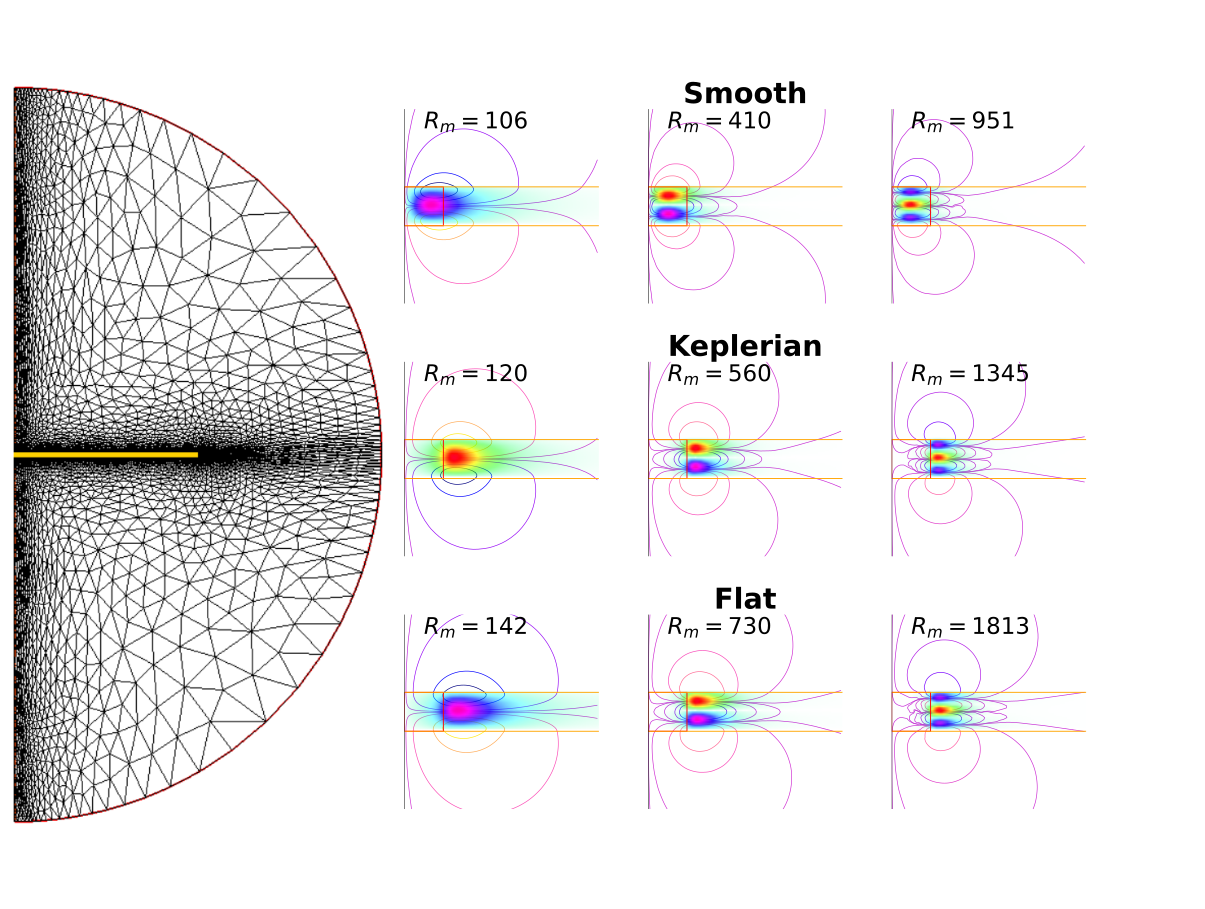}
    \centering
    \caption{(a) Left: Meshing domain for numerical simulations. (b) Right : Isovalues of the poloidal and toroidal components of the magnetic field for Smooth, Keplerian and Flat azimuthal velocity profiles at different $R_m$.  The axis of rotation is represented by the vertical black line. The radial limit of the bulge is represented by the vertical red line. The conducting domain lies between the two horizontal orange lines. Increasing $R_m$ we see that the magnetic mode evolves from quadrupolar (left), octopolar (middle), and to a higher mode (right).
    }  
    \label{fig:Mesh}
\end{figure}

The problem is axisymmetric. In the $(r,z)$-plane the galaxy is modeled by a rectangle of dimensions $h~\times~R$, and the computational domain  is a semi-disc of radius $R'$ (see Fig. \ref{fig:Mesh}a). The electrical conductivity tensor is defined by (\ref{eq:conductivity permeability tensors}) in the rectangle, and zero elsewhere.
The corresponding boundary conditions are
\begin{equation}
    T(r=0)=P(r=0)=T(z=\pm\frac{h}{2},r\le R)=0,
\end{equation}
the last condition insuring that at the border of the conducting domain, the normal component of the current density vanishes. 
For $\rho=R'$ with $\rho=\sqrt{r^2+z^2}$, we impose
\begin{equation}
T(\rho=R')=\left(\frac{\partial P}{\partial r}-\frac{P}{R'}\right)(r=R')=0,
\end{equation}
meaning that the magnetic field matches with a dipolar magnetic field at the limit of the computational domain $\rho=R'$.

%%%%%%%%%%%%%%%%%%%%%%%%%%%%%%%%%%%%%%%%%%%%%
\subsection{Growth rates}
%%%%%%%%%%%%%%%%%%%%%%%%%%%%%%%%%%%%%%%%%%%%%
\begin{figure}
    \centering
    \includegraphics[width = \linewidth]{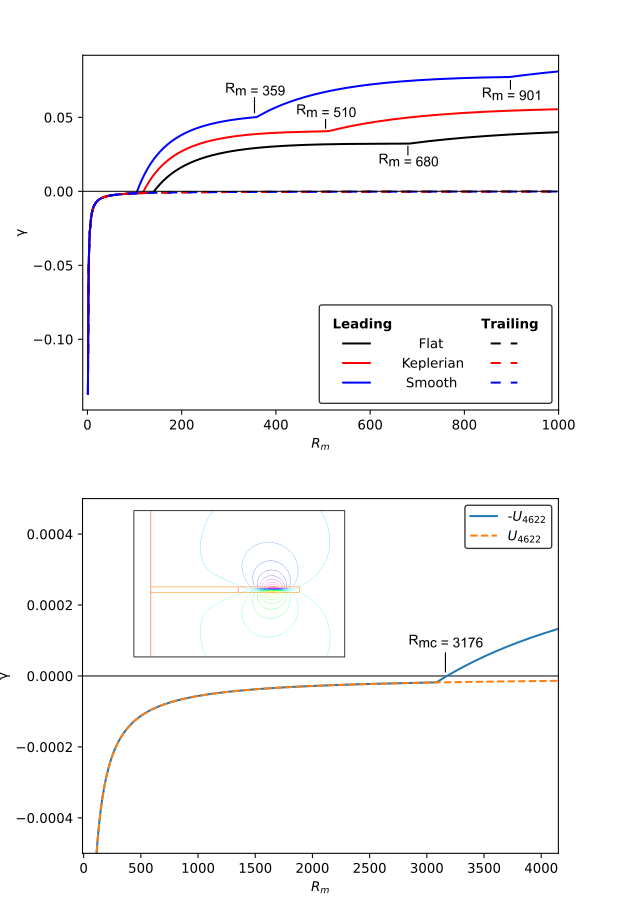}
    \caption{(a) Top : Growth rate $\gamma$ versus $R_m$ for $U_K, U_F, U_S$ (dashed curves) and $-U_K, -U_F, -U_S$ (solid curves). (b) Bottom: Growth rate $\gamma$ versus $R_m$ for $U_{4622}$ (dashed curve) and $-U_{4622}$ (solid curves).  In the inset, the isovalues of the poloidal component of the magnetic field are plotted for $R_m = 3176$.}

    \label{fig:growthRate}
\end{figure}
In Fig. \ref{fig:growthRate}, the magnetic growthrate $\gamma$ is plotted versus $R_m$ for each velocity profiles (\ref{eq:UK}-\ref{eq:U4622}).
As predicted in sections  \ref{sec:galaxies} and \ref{sec:NGC 4622}, dynamo ($\gamma>0$) is only possible  if we replace $U_S$, $U_K$, $U_F$ and $U_{4622}$ by their opposites (or if we replace the pitch angle by its opposite).
For $U_S$, $U_K$, $U_F$ it means that only the leading arms configuration can lead to dynamo action. For $U_{4622}$, only the configuration with a leading arm surrounded by trailing arms can lead to a dynamo. Increasing the value of $\sigma_\perp / \sigma_\parallel$ does not change this qualitative conclusion (as illustrated in Fig. \ref{fig:GrVSRm} of Appendix for $U=U_S$).

From Fig. \ref{fig:growthRate}, we note that the dynamo threshold $R_m^c$, above which $\gamma > 0$, is much higher for $-U_{4622}$ ($R_m^c=3176$) than for $-U_S$, $-U_K$ and $-U_F$ ($R_m^c$=106, 120 and 142).  Also the value of $\gamma$ is smaller by about two orders of magnitude, for $-U_{4622}$.
This agrees well with (\ref{condition}) as, for NGC 4622, $p$ and $\Omega'$ are about 10 times smaller than in the three other cases.

For $-U_S$, $-U_K$ and $-U_F$ there are magnetic mode transitions, from quadrupolar to octopolar at $R_m = 359, 510$ and 680 respectively, and from octopolar to a higher degree at $R_m = 901, 1295$ and 1763 respectively.  For $-U_{4622}$, the magnetic eigenmode is quadripolar and generated in the outer part of the conducting domain, corresponding to the leading arms (Fig. \ref{fig:growthRate}b). A 3D representation of the magnetic field lines is shown in Appendix (Fig. \ref{fig:Model3D}) for $U=-U_S$ and $U=-U_{4622}$.

\subsection{Characteristic time}
In the high-$R_m$ limit, and in all four cases $-U_S, -U_K, -U_F$ and $-U_{4622}$, the growthrate does not decrease toward zero, which is the signature of fast dynamo action (Fig. \ref{fig:growthRate}). As shown in \cite{plunian_fast_2022}, the dynamo takes place (i) at a characteristic radius where the shear is maximum, located here at $r\sim r_b$, and (ii) with a characteristic time which is the turnover time at that scale, here $\tau \sim |\Omega(r_b)^{-1}|$.  
  With a characteristic velocity  $U\sim 10^2$~km~s$^{-1}$ \citep{persic1996universal} and a characteristic bulge's radius $r_b\sim1 - 10$~kpc, we find that spiral-arm galaxies have  a characteristic turnover time
$\tau\sim10^7-10^8$~years which is a fraction of spiral-arm galaxy's age.

\subsection{Characteristic scale}
We note that the larger the $\Omega$, the smaller the vertical magnetic structures (Fig. \ref{fig:Mesh}), in agreement with the fast dynamo model of \cite{plunian_fast_2022}. The characteristic scale of magnetic structures is of order $R_m^{-1/2}r_b$. At this scale the magnetic diffusion time is equal to the turnover time.

\section{Conclusion}
In this paper, we studied the question of whether a dynamo effect in a spiral-arm galaxy can be a simple consequence of the existence of its spiral arms. The answer could be yes, provided the arms would be leading and not trailing as usually observed in galaxies, or provided the shear would increase with $r$ ($|\Omega|'>0$), which is usually also not observed in galaxies. Having both leading arms and an increasing shear with $r$ would not lead to dynamo either. Therefore, based on current observations, we do not find any possibility to have a galactic magnetic field generated only by a simple differential rotation, without requiring an effect of turbulence e.g. an alpha-effect. This does not mean that the geometry of the arms plays no role. In fact, it remains true that electric currents flow preferentially along the spiral arms of the galaxy, corresponding to an anisotropic electrical conductivity effect. Therefore a complete model of a galactic dynamo should include at least the three effects: differential rotation, effect of turbulence like an alpha-effect, and an anisotropic effect. 
We can also think of large-scale fluctuations in azimuthal velocity, so that the sign of $\Omega'$ is modified at certain points. It would then be possible for the anisotropic dynamo to take hold. In NGC 4622, the sign of $\Omega'$ is indeed changed at $r=30$ arcsec. However, it is accompanied by the transformation of the trailing arms into leading arms, which again rules out the possibility of an anisotropic dynamo. From these observations, we can suggest that the anisotropic dynamo effect induced by shear leads to a selection principle: if the condition (7) applied, then the fast dynamo would produce a strong magnetic field, inducing strong Lorentz forces that would reorganize the differential rotation and/or pitch angle until condition (7) is no longer satisfied. This remains to be demonstrated in more detailed dynamical numerical simulations.
\section*{Acknowledgement}
\begin{acknowledgments}
We acknowledge the support of the Programme et Équipements Prioritaire pour la Recherche (PEPR Origins) for its financial support.
Numerical simulations were carried out using Freefem++ \citep{MR3043640} and 3D modelling was done using Paraview software \citep{ahrens200536}.%.Acknowledge people, funding sources \Fp{PNP ? PEPR ?}, and facilities here. For example, "We acknowledge the support of NASA under grant number XYZ."
\end{acknowledgments}

\appendix
%\section{Electrical conductivity ratio}
\label{appendix:A}

\begin{figure}%[h]
    \centering
    \includegraphics[width = 0.8\linewidth]{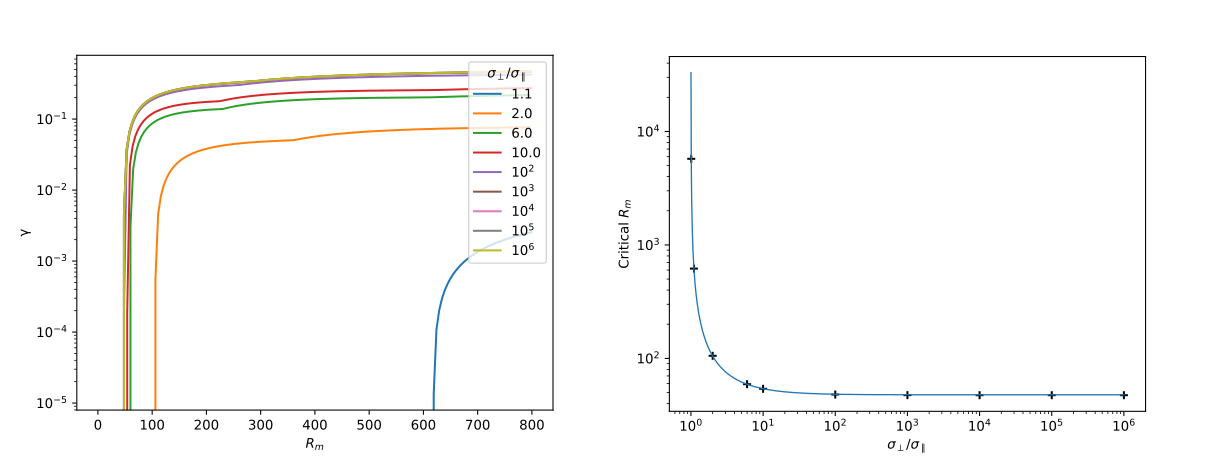}
    \caption{For $U=-U_S$ and different values of $\sigma_\perp/\sigma_\parallel$, (a) Left: the growth rate is plotted versus $R_m$. (b) Right : The critical $R_m$ above which dynamo occurs, is plotted versus $\sigma_\perp/\sigma_\parallel$.}
    \label{fig:GrVSRm}
\end{figure}

 In Fig. \ref{fig:GrVSRm}a, the growthrate is plotted versus $R_m$ for $U=-U_S$ and several values of $\sigma_\perp / \sigma_\parallel > 1$. In Fig. \ref{fig:GrVSRm}b, the critical magnetic Reynold number is plotted versus $\sigma_\perp / \sigma_\parallel$ .

\begin{figure}
    \centering
    \includegraphics[width = 0.8\linewidth]{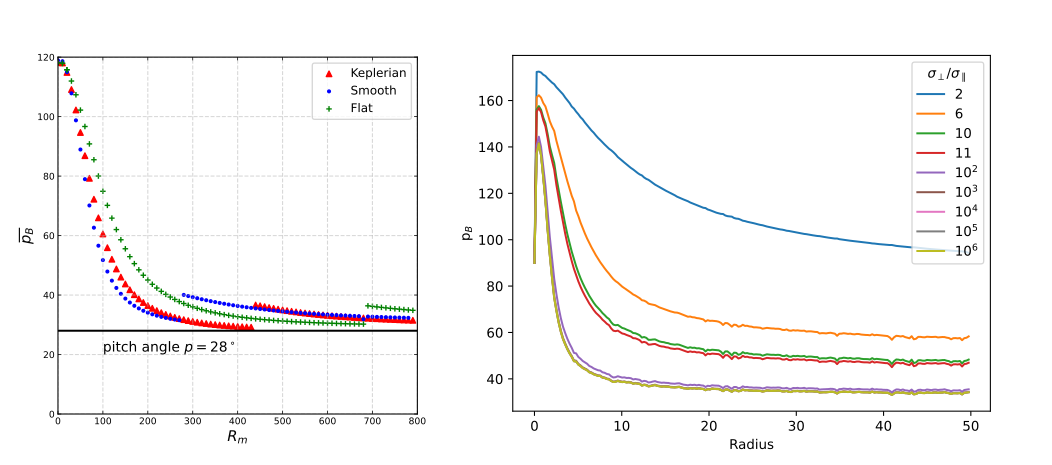}
    \caption{(a) Left : The radial mean of the magnetic pitch angle $\bar{p_B}$ is plotted versus $\Rm$, for $U= -U_K, -U_S,-U_F$, and $\sigma_\perp / \sigma_\parallel = 10^6$. (b) Right : For $U=-U_S$ and $R_m = 200$, the magnetic pitch angle $p_B$ is plotted versus $r$ for different values of  $\sigma_\perp / \sigma_\parallel$.}

    \label{fig:MagneticPitch}
\end{figure}

The magnetic pitch angle $p_B$, is defined by 
\begin{equation}
 p_B = \arctan\left(\frac{B_r}{B_\theta}\right).
\end{equation}
In Fig. \ref{fig:MagneticPitch}a, the radially averaged magnetic pitch angle $\bar{p_B}$ is plotted versus $\Rm$, for $U= -U_K, -U_S,-U_F$, and $\sigma_\perp / \sigma_\parallel = 10^6$.
In the limit of large $\Rm$, $\bar{p_B}$ approaches asymptotically the motion pitch angle $p$, as already observed in spiral-arms galaxies \citep{van2015magnetic, beck2016magnetic}.
In Fig. \ref{fig:MagneticPitch}b, for $U=-U_S$ and $R_m = 200$, the magnetic pitch angle $p_B$ is plotted versus $r$ for different values of  $\sigma_\perp / \sigma_\parallel$. 

\begin{figure}
    \centering
    \includegraphics[width = \linewidth]{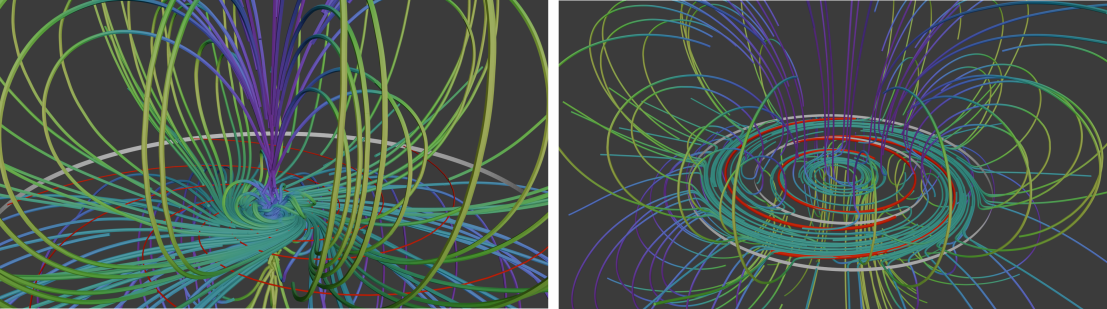}
    \caption{ A 3D representation of the magnetic field lines for (a) Left: $U=-U_S$ and $R_m=350$, (b) Right: $U=-U_{4622}$ and $R_m=3180$. The green and blue colors correspond to opposite signs of $\bB_z$. The red lines are perpendicular to $\bq$, so in the direction of the spiral arms, assuming axisymmetry. In both figures, the outer grey circular ribbon  marks the radial edge of the conducting domain. In the right figure, the inner grey circular ribbons marks the transition between the leading and trailing arms. }

    %The normalized $B_z$ is represented in color, red when $B_z$ is positive and blue when it is negative. The parameters used corresponds to the case (b) in Fig.\ref{fig:spiral}
    \label{fig:Model3D}
\end{figure}

In Fig. \ref{fig:Model3D}, a 3D representation of the magnetic field is plotted for (a) $U=-U_S$ (respectively (b) $U=-U_{4622}$)  and $R_m=350$ 
(respectively $R_m=3180$).

\newpage

\bibliography{biblio}{}
\bibliographystyle{aasjournal}

\end{document}